\documentclass{emulateapj}

\usepackage[]{natbib}
\usepackage{color}

\begin{document}

\title{High-Resolution Near Infrared Spectroscopy of HD~100546: I. Analysis of asymmetric ro-vibrational OH emission lines}

\author{Joseph P. Liskowsky \& Sean D. Brittain}
\affil{Department of Physics \& Astronomy, 118 Kinard Laboratory, Clemson University, Clemson, SC 29634, USA; sbritt@clemson.edu}

\author{Joan R. Najita}
\affil{National Optical Astronomy Observatory, 950 N. Cherry Ave., Tucson, AZ 85719, USA; najita@noao.edu}

\author{John S. Carr}
\affil{Naval Research Laboratory, Code 7211, Washington, DC 20375, USA; carr@nrl.navy.mil}

\author{Greg W. Doppmann}
\affil{W.M. Keck Observatory, 65-1120 Mamalahoa Hwy, Kamuela, HI  96743, USA, gdoppmann@keck.hawaii.edu}

\author{Matthew R. Troutman} 
\affil{Department of Physics \& Astronomy, University of Missouri - St. Louis, St. Louis, MO 63121, USA; troutmanm@umsl.edu}

\begin{abstract}
We present observations of ro-vibrational OH and CO emission from the Herbig Be star HD~100546. The emission from both molecules arises from the inner region of the disk extending from approximately 13 AU from the central star. The velocity profiles of the OH lines are narrower than the velocity profile of the [\ion{O}{1}]~$\lambda$6300\AA\   line indicating that the OH in the disk is not cospatial with the \ion{O}{1}. This suggests that the inner optically thin region of the disk is largely devoid of molecular gas. Unlike the ro-vibrational CO emission lines, the OH lines are highly asymmetric. We show that the average CO and average OH line profiles can be fit with a model of a disk comprised of an eccentric inner wall and a circular outer disk.  In this model, the vast majority of the OH flux ($75\%$) originates from the inner wall, while the vast majority of the CO flux ($65\%$) originates on the surface of the disk at radii greater than 13~AU.  Eccentric inner disks are predicted by hydrodynamic simulations of circumstellar disks containing an embedded giant planet.   We discuss the implications of such a disk geometry in light of models of planet disk tidal interactions and propose alternate explanations for the origin of the asymmetry. 
\end{abstract}

\keywords{stars: circumstellar matter, individual (HD~100546), protoplanetary disks}

\section{INTRODUCTION}
\label{sec:intro}

An important tool for tracing the evolution of circumstellar disks is the measurement of the spectral energy distribution (SED) of young stars \citep[e.g.][]{1989AJ.....97.1451S, 1992ApJ...397..613H, 1998A&A...331..211M}. The differences in the SEDs of young stars have led some to propose that they reflect an evolutionary sequence from optically thick disks to transitional disks (i.e. disks with optically thick outer disks and optically thin inner disks) to optically thin disks.  A popular interpretation of transitional objects is that they reflect ongoing planet formation, but there are other processes that can also give rise to disks with transitional SEDs.

The proposed scenarios for the origin of transitional disks can be distinguished by the radial distribution of gas in the disk \citep{2007MNRAS.378..369N}.  The grain-growth and planetesimal formation scenarios predict an optically thin, gas-rich inner disk \citep[e.g.][]{1989AJ.....97.1451S}. A system that is forming a Jovian mass planet will show a radial gap in the gaseous inner disk in the vicinity of the planet's orbit \citep[e.g.][]{skrutskie1990}.    Systems that have formed massive companions \citep[$>$5~M$\rm_J$; e.g. ][]{1999ApJ...526.1001L} or are being photoevaporated away  \citep[e.g.][]{Clarke2001} will be depleted in gas in the optically thin region of the disk.  Thus, we can determine which scenario likely applies to any given transition object by probing the distribution of gas in the inner disk. As the planet-forming region of disks is generally spatially unresolved, high-resolution spectroscopy becomes a surrogate by spectrally resolving the velocity of gas. Assuming Keplerian rotation, spatial information can be extracted if the stellar mass and disk inclination are known.

Work to date that takes this approach presents conflicting results for well-known intermediate-mass transition objects such as HD~100546. HD~100546 is a nearby \citep[d=103$^{+7}_{-6}$ pc; ][]{1997A&A...324.33V} Herbig Be star  with an ostensibly massive disk \citep[$\rm M_{disk}=0.072~M_\odot$; ][]{1998A&A...336..565H}. Modeling of the SED  \citep{2003A&A...401..577B} and imagery with the {\it Hubble Space Telescope} \citep{2005ApJ...620...470G} indicate the presence of an inner hole in the dust distribution at 10-13~AU. While ro-vibrational CO observations of HD~100546 show that there is no CO within 13~AU of the star \citep{2009ApJ...702...85B,2009A&A...500.1137V}, evidence of ongoing accretion indicates some gas persists in the inner hole \citep{2006A&A...457..581G}.

In addition, the spectral profile of the [\ion{O}{1}] $\lambda$6300\AA\ line indicates that some gas extends inward to at least 0.8~AU \citep{2006A&A...449..267A}.  Following the work by \citet[][2000]{1998ApJ...502L..71S},  \citet{2006A&A...449..267A} suggest that the \ion{O}{1} emission may arise from the photodissociation of OH. If this is the case, then the inner 13~AU of the disk is curiously rich in OH while depleted in CO, contrary to predictions for the chemistry of inner disks \citep[e.g.][]{thi05}.

Rotational OH emission has been detected from the disk surrounding HD~100546, however, the spatial and spectral resolution are too low to resolve the line and thus pinpoint the location of the gas within the disk \citep{2010A&A...518L.129S}. Higher excitation ro-vibrational OH emission has been reported from other young stars: from the classical T Tauri stars AS 205A and DR Tau \citep{Sal08}, as well as SVS 13 \citep{naj07} and V1331 Cyg \citep{naj07,2011ApJ...738..112D}, and several Herbig Ae/Be stars \citep{Mandell08, Fedele2011}. \citet{Mandell08} report P2.5 through P9.5 ro-vibrational lines of OH for two objects:  AB Aur and MWC 758. 

In this paper we report the observation of ro-vibrational OH and CO emission from the transitional disk surrounding HD~100546 (Sec. \ref{sec:Results}). We compare the line profile of the OH emission to that of the ro-vibrational CO emission and the [\ion{O}{1}] $\lambda$6300\AA\ emission line in order to explore the properties in the region of the disk from which these emission features arise (Sec. \ref{sec:Modeling}).  A detailed comparison of multi-epoch CO observations of this source is presented in a companion paper \citep{2012} (hereafter Paper II).

\section{Observations}
\label{sec:Observations}
\begin{figure}
    \begin{center}
	\includegraphics[scale=.7]{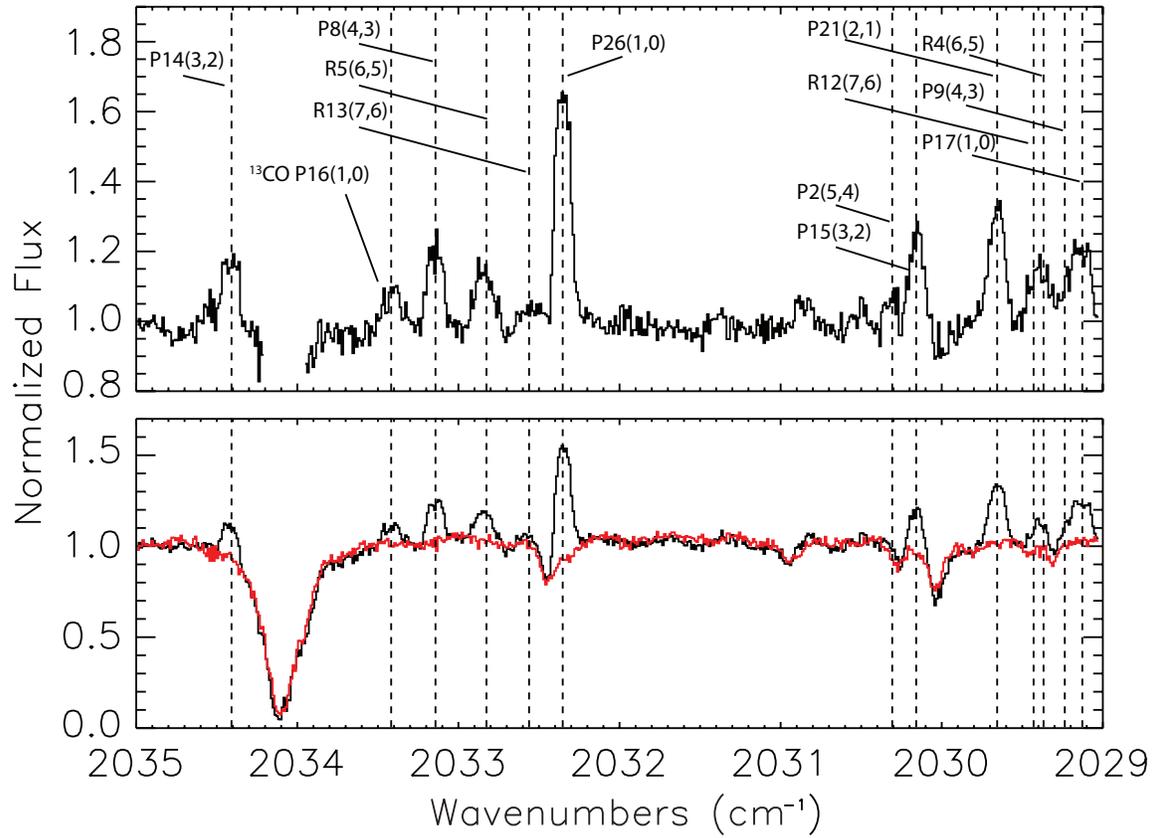}
	\caption[M-band spectra of HD100546]{High resolution near infrared spectra of HD~100546 in the observed frame. The top panel shows the ratioed $M-$band spectra containing the ro-vibrational CO transitions. Numerous emission features are evident and are labeled.  The positions of these features in this spectrum are marked by vertical dot-dashed lines.  The bottom panel shows the observed spectrum for this wavelength range and the telluric standard in red.  The strong emission visible in the top panel is also clearly seen here. The gaps in the spectrum are of regions where the transmittance of the atmosphere is less than 50\%. The fundamental ro-vibrational CO P26 line and numerous hot band features are evident.}
	\label{fig:obs1}
   \end{center}
\end{figure}

\begin{figure}
    \begin{center}
	\includegraphics[scale=.7]{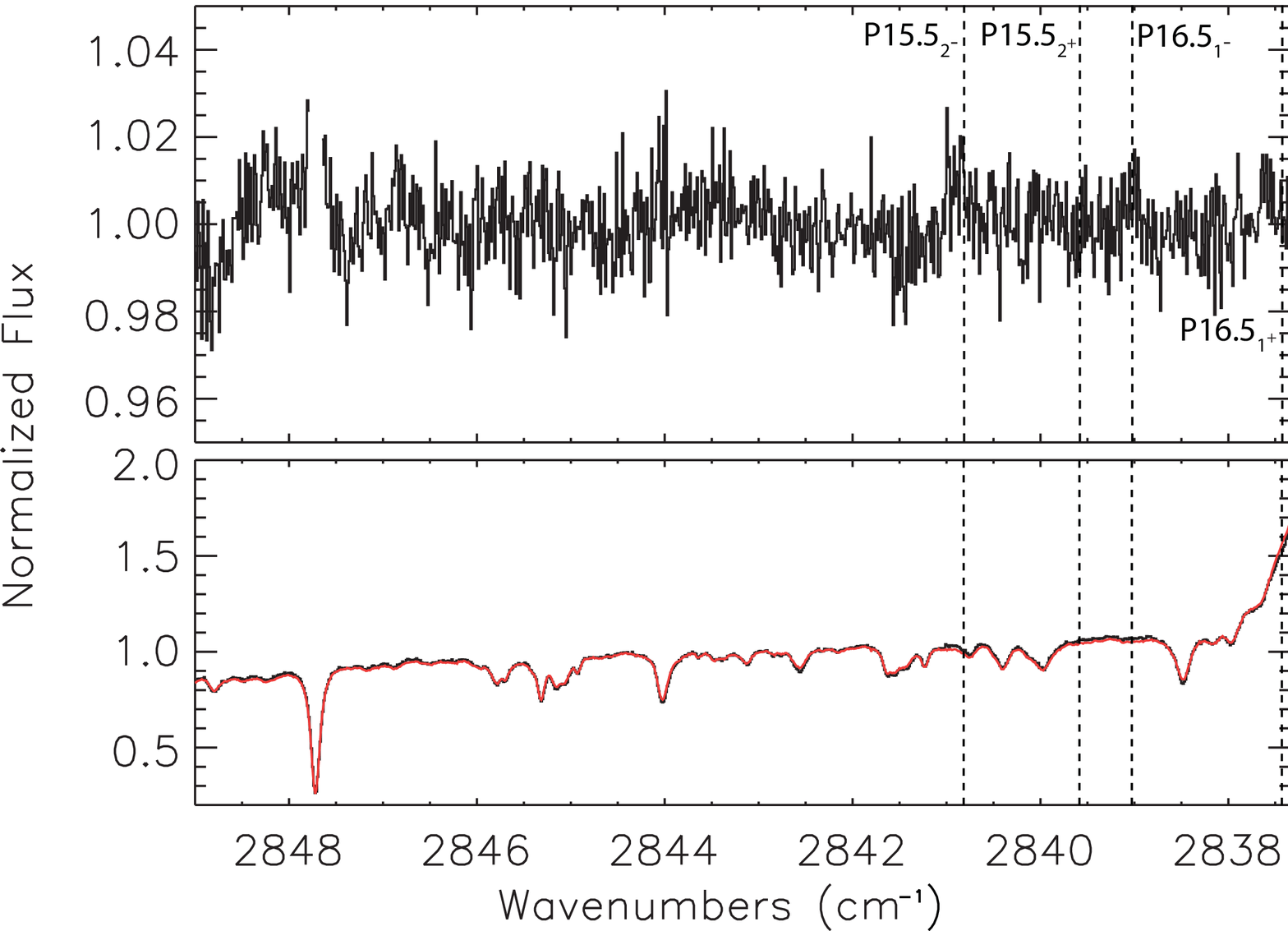}
	\caption[L-band spectra of HD100546]{High resolution near infrared spectra of HD~100546 in the observed frame. The top panel shows the $L-$band spectra containing the ro-vibrational OH transitions while the bottom panel presents the observed spectrum (black) and the telluric standard (red). The gaps in the spectrum are of regions where the transmittance of the atmosphere is less than 50\%. The positions of the v=1-0 OH lines are labeled with vertical dot-dashed lines. The  P15.5$_{2\pm}$ and P16.5$_{1\pm}$ ro-vibrational emission lines are not detected. We use the noise in the continuum to set an upper limit on the equivalent widths of these lines, which are then used to put an upper limit on the rotational temperature.  Detections of emission lines in this range are below the 1-$\sigma$ level.}
	\label{fig:obs2}
   \end{center}
\end{figure}

\begin{figure}
    \begin{center}
	\includegraphics[scale=.7]{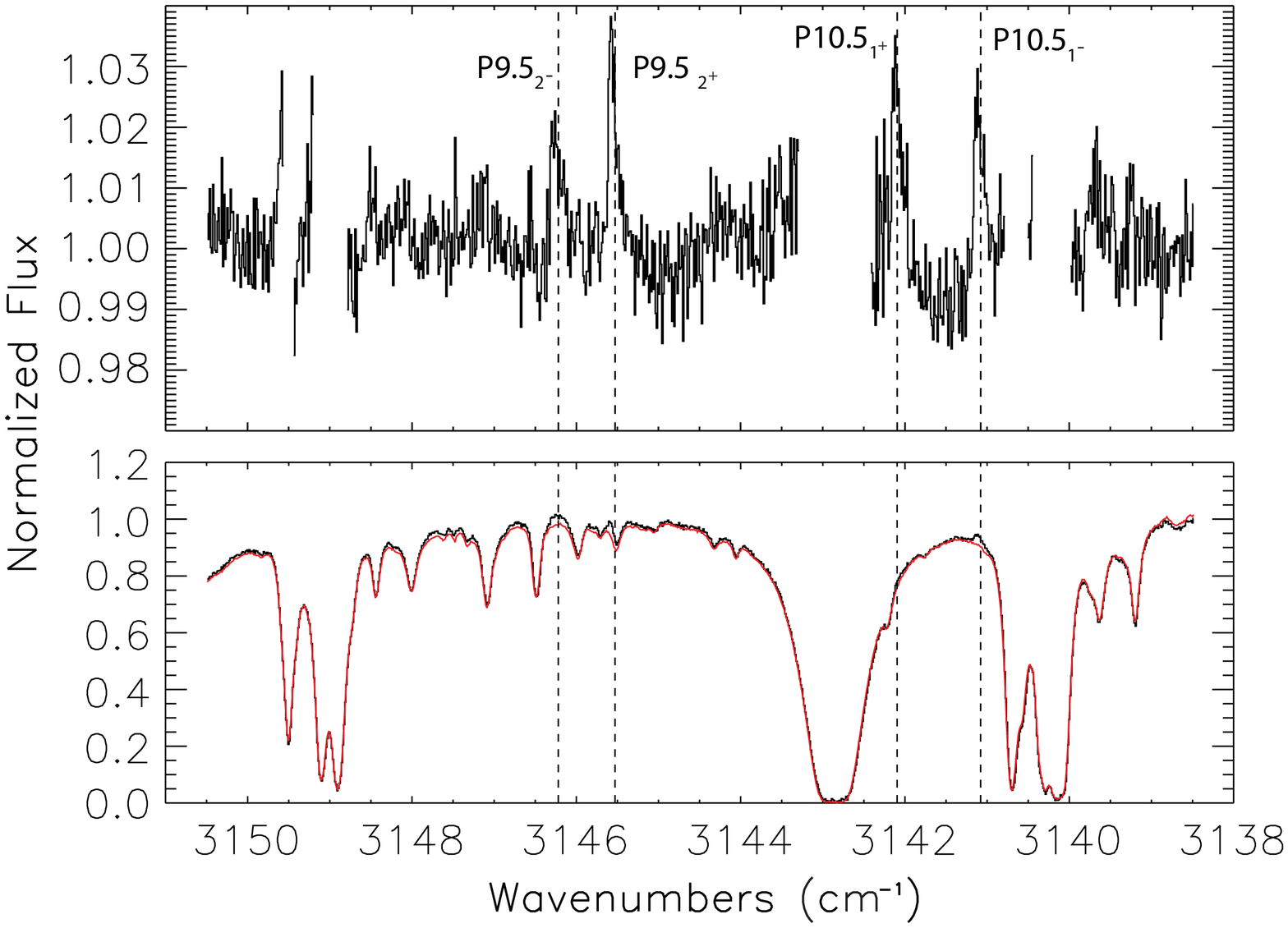}
		\caption[L-band spectra of HD100546]{High resolution near infrared spectra of HD~100546 in the observed frame. The top panel shows the ratioed $L-$band spectrum containing the ro-vibrational OH transitions while the bottom panel presents the observed spectrum (black) and the telluric standard (red). The gaps in the spectrum are of regions where the transmittance of the atmosphere is less than 50\%. The positions of the v=1-0 OH lines are labeled with vertical dot-dashed lines. The P9.5$_{2\pm}$ and P10.5$_{1\pm}$ ro-vibrational emission lines are detected and labeled with dot-dashed vertical lines.}
	\label{fig:obs3}
   \end{center}
\end{figure}

High-resolution (R=50,000), near-infrared spectra of HD~100546 were acquired December 22, 
2010 and December 23, 2010 using PHOENIX at the Gemini South telescope with the four pixel (0$\arcsec$.34) slit 
 \citep{2003SPIE.4834..353H,2000SPIE.4008..720H,1998SPIE.3354..810H}. A summary of observations is presented in Table \ref{tab:observations}. The position angle of the slit was 90$\degr$ East of North for all observations. The spectra are centered at 3145 cm$^{-1}$, 2844 cm$^{-1}$, and 2032 cm$^{-1}$ (Figs. \ref{fig:obs1}, \ref{fig:obs2}, \ref{fig:obs3}).  The PSF of the continuum is  $0.\arcsec7$.    The spectral settings were selected to minimize telluric absorption of the CO and OH emission lines. The $M-$band observation includes the v=1-0 P25, P26, and P27 lines of CO as well as numerous other hot band transitions ($\Delta$ v=1, v$^{\prime} \geq 2$).  See \citet{2009ApJ...702...85B} for complete line identifications. The $L-$band observations cover the P9.5$_{2^{\pm}}$, P10.5$_{1^{\pm}}$, P15.5$_{2^{\pm}}$, and P16.5$_{1^{\pm}}$ v=1-0 lines of OH.  The notation is described in the appendix.
  
Because of the thermal background in the $L-$ and $M-$ bands, the data were observed in an ABBA sequence where the telescope was nodded 5$\arcsec$ between the $A$ and $B$ positions. The observations were combined as $(A-B-B+A)/2$ in order to cancel the sky emission to first order. Each frame was flat fielded and scrubbed for hot pixels and cosmic ray hits. The sky lines were tilted by about a pixel along the full length of the slit. The data were rectified in the spatial direction by fitting a polynomial to the centroid of the point spread function (PSF) of the continuum. The data were rectified in the spectral direction by fitting a sky emission model to each row and interpolating the wavelength solution of each row to the row in the middle of the detector. To calibrate the spectrum, we fit a telluric model to the spectrum. The model was generated by the Spectral Synthesis Program  \citep{1974JQSRT..14..803K}, which accesses the 2000HITRAN molecular database \citep{2003JQSRT..82....5R}. The fit to strong lines is generally accurate to within a fraction of a pixel ($\sim$1~km~s$^{-1}$). However, near saturated lines (particularly water lines), there is a degeneracy between the width of the line and the dispersion; the latter varies across the array. The level of uncertainty in the calibration near such saturated water lines is typically at the level of a pixel ($\sim$1.5~km~s$^{-1}$ for Phoenix).  Of the four OH lines we observed, one line, P10.5$_{1^{+}}$, was on the wing of a broad, saturated telluric water line. We shifted this line by 1 pixel blueward relative to the other three lines such that the standard deviation of the residuals between this line and the other three OH lines corresponded to the standard deviation of the noise along the continuum. 

We extracted the one dimensional spectrum of the star weighted by the stellar PSF. The extraction window was 13 rows (1$\arcsec$.1). The spectrum of HD~100546 was ratioed with the spectrum of the standard star. Regions of the spectrum where the transmittance falls below 50\% are omitted (Figs. \ref{fig:obs1}, \ref{fig:obs2}, \ref{fig:obs3}). The equivalent widths of the OH lines are reported in Table 2, and a summary of the CO line positions is reported in Table 3.  

\section{Results}
\label{sec:Results}
\subsection{OH Temperature}
\begin{figure}
    \begin{center}
	\includegraphics[scale=.7]{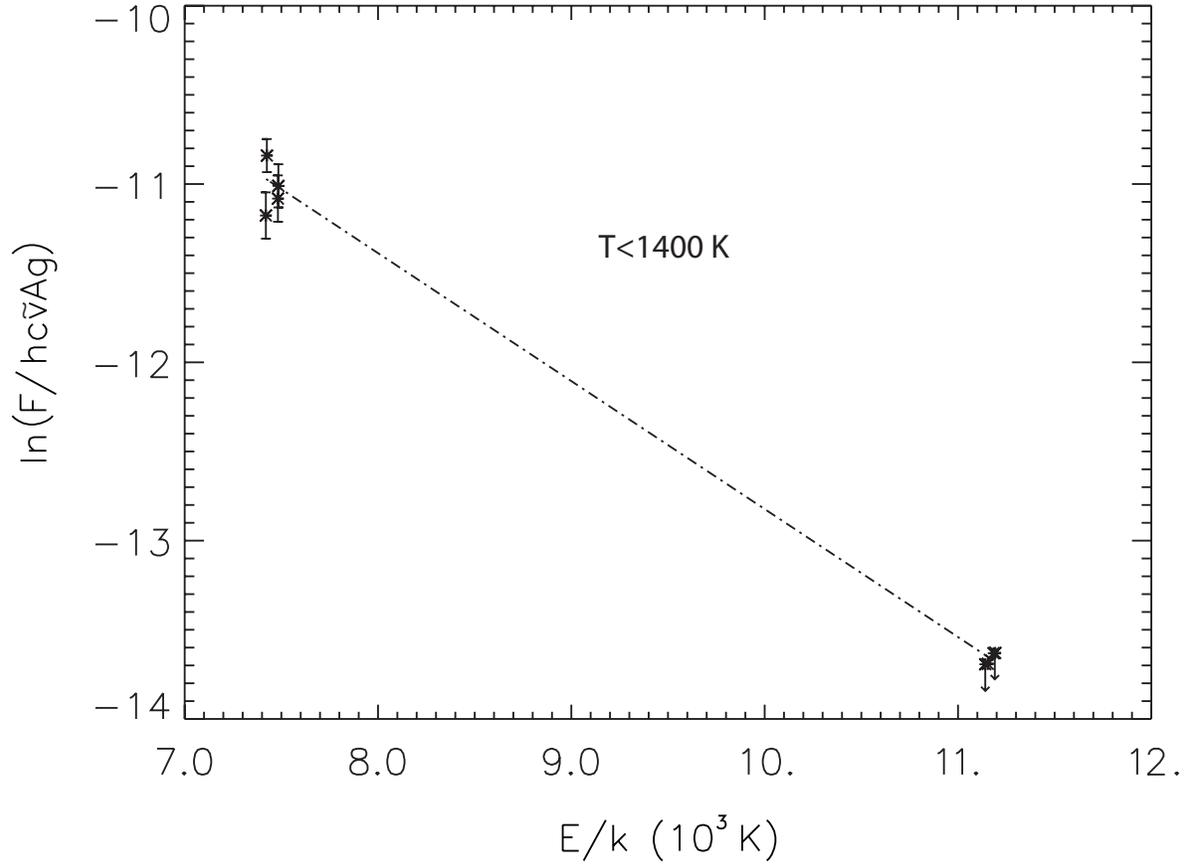}
	
	\caption[Exc Diag]{Excitation diagram of ro-vibrational OH emission lines.  The quantity ln(F/hc$\tilde{\nu}$Ag) is plotted against E/k, the energy of the upper level, where F is the line flux (in erg/s/cm$^{2}$/cm$^{-1}$), $\tilde{\nu}$ is the wavenumber of the transition (in cm$^{-1}$), g is the multiplicity of the upper level, and A is the Einstein A coefficient of the transition. Thus the rotational temperature of the gas is given by the negative reciprocal of the slope. The solid line shows the fit to the detected values and the upper limits.  The 1$\sigma$ upper limit on the rotational temperature of the OH gas, based on the best fit line, is 1400~K.}
	
	\label{fig:obs4}
   \end{center}
\end{figure}

We detect the v=1-0 P9.5$_{2^{\pm}}$, P10.5$_{1^{\pm}}$ OH lines (Fig. \ref{fig:obs3}) and not the v=1-0 15.5$_{2^{\pm}}$, P16.5$_{1^{\pm}}$ lines (Fig. \ref{fig:obs2}). We use the higher rotational lines to place an upper limit on the rotational temperature of the gas. If the gas is optically thin, the flux of an emission line is related to the column density of molecules in the upper energy state,

\begin{equation}
F_{ij}=\frac{hc \tilde{\nu_{ij}} A_{ij}N_{i}}{4 \pi d^{2}}
\end{equation}

 \noindent where $N_{i}$ is the number of OH molecules in the upper energy state, $hc\tilde{\nu_{ij}}$ is the energy of the emitted photon, and $A_{ij}$ is the Einstein A coefficient. For gas in local thermodynamic equilibrium, the relative population of the energy levels is given by the Boltzmann distribution,

\begin{equation}
N_{i}  = Ng_{i}e^{-E_{i}/kT}/Q
\end{equation}

\noindent where $g_{i}$ is the statistical weight of the upper energy state, $E_{i}$ is the energy of the upper state, N is the total number of molecules, and $Q$ is the partition function.  Thus $ln(F_{ij}/hc \tilde{\nu_{ij}} A_{ij}g_{i}) = -E_{i}/kT + ln(N / 4\pi d^2 Q) $ so that the rotational temperature is given by the negative reciprocal of the slope of a linear least squares fit of $ln(F_{ij}/hc \tilde{\nu_{ij}} A_{ij}g_{i})$  and $E_{i}/kT$. Since we only detect the P9.5$_{2^{\pm}}$, P10.5$_{1^{\pm}}$ lines we can only place an upper limit on the rotational temperature of the gas. We fit the detected lines and the 1$\sigma$ upper limit of the non-detected lines and find that T$\rm_{rot} < $1400K (Fig. \ref{fig:obs4}). This upper limit is consistent with the range of CO rotational temperatures at the inner edge of the disk \citep[T = 1400~$\pm$400~K;][]{2009ApJ...702...85B}.  As the critical densities for rotational excitation of the levels we observe are high, it is possible that the OH is not rotationally thermalized. However, it seems likely that a disk environment that is dense enough to thermally populate the $v=1$ vibrational state of CO \citep{2009ApJ...702...85B} would be dense enough to populate the rotational states of OH. 

\subsection{Comparison of HWZI of CO, OH, and OI}  
We previously reported the January 2006 detection of ro-vibrational CO emission from HD~100546 \citep{2009ApJ...702...85B}. In our earlier report, we showed that the gas is excited by a combination of UV fluorescence and collisional excitation and that the emitting region extends from 13~AU to at least 100~AU. In data acquired five years later, which are presented here, we find that the equivalent width of the fundamental CO emission lines increased by 50\% whereas the average equivalent width of the hot band lines increased by 30\% (Paper II). We also find that while the line profile of the hot band CO emission lines remained constant, the line profile of the v=1-0 emission lines varied. In the present epoch, the v=1-0 CO P26 line profile is symmetric, whereas the majority of the luminosity in the hot band CO lines is on the blue side of the lines.

A detailed study of multi-epoch CO observations is presented in Paper II. In that paper we show that over three epochs spanning 2003-2010 the spectral line profile and equivalent widths of the v=1-0 CO lines vary;  the equivalent widths of the hot band lines are less variable and their spectral line profile remains constant.  Because variations in the M-band continuum can account for the modest variations in the equivalent width of the hot band lines, {} we assume that the hot band lines are approximately constant in flux, and that the v=1-0 lines vary more dramatically. In this paper we restrict our comparison of the OH and CO line profiles to data taken contemporaneously, and we focus on the hot band CO lines because they are less variable. 

\begin{figure}
    \begin{center}
	\includegraphics[scale=1.]{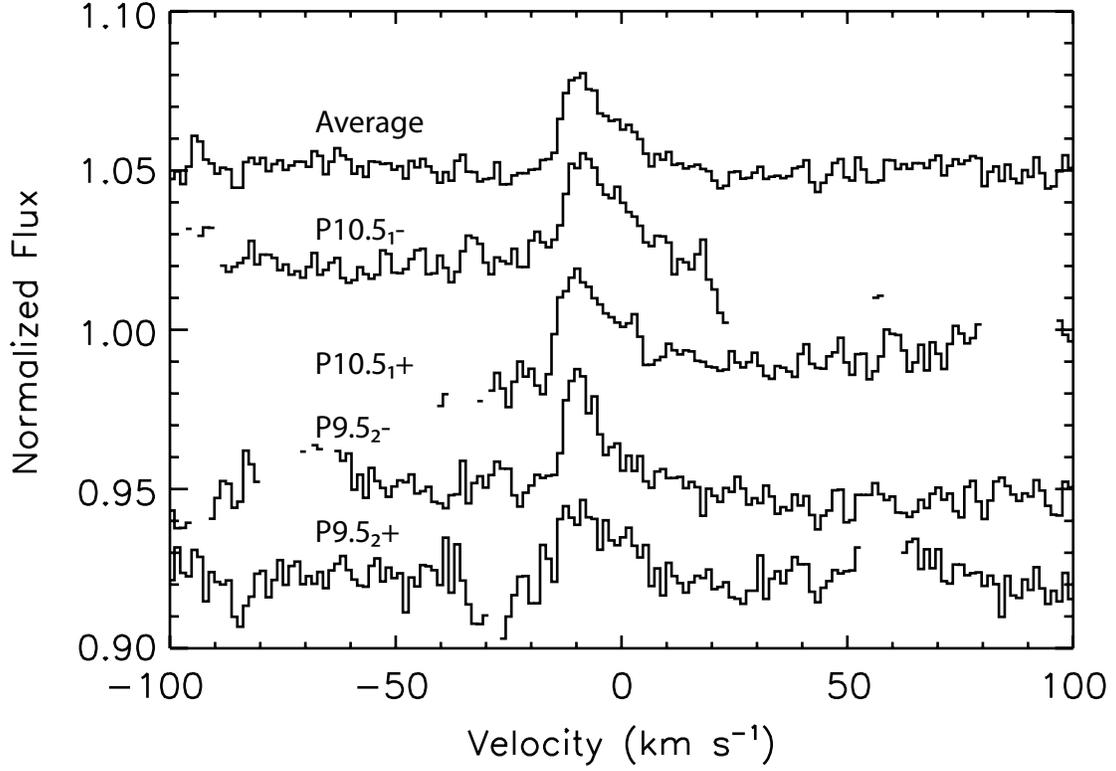}
	\caption[OH lines]{Individual OH lines used to construct the average OH line profile in the rest frame of the star.  The S/N of the average continuum is $\sim$325.  The average OH line to continuum ratio is $\sim3\%$ and the continuum noise is $\sim0.6\%$. The HWZI is 12~$\pm$2 km~s$^{-1}$. The asymmetric line shape is evident in most of the individual lines, with approximately 80$\%$ of the emission in the blue and 20$\%$ in the red. The P10.5$_{1^{+}}$ line was shifted 1.5~km~s$^{-1}$ blueward to line up with the other OH lines such that residuals between the P10.5$_{1^{+}}$ line and the other OH lines corresponded to the noise along the noise along the continuum. }  
	\label{fig:obs5}
   \end{center}
\end{figure}

\begin{figure}
    \begin{center}
	\includegraphics[scale=1.]{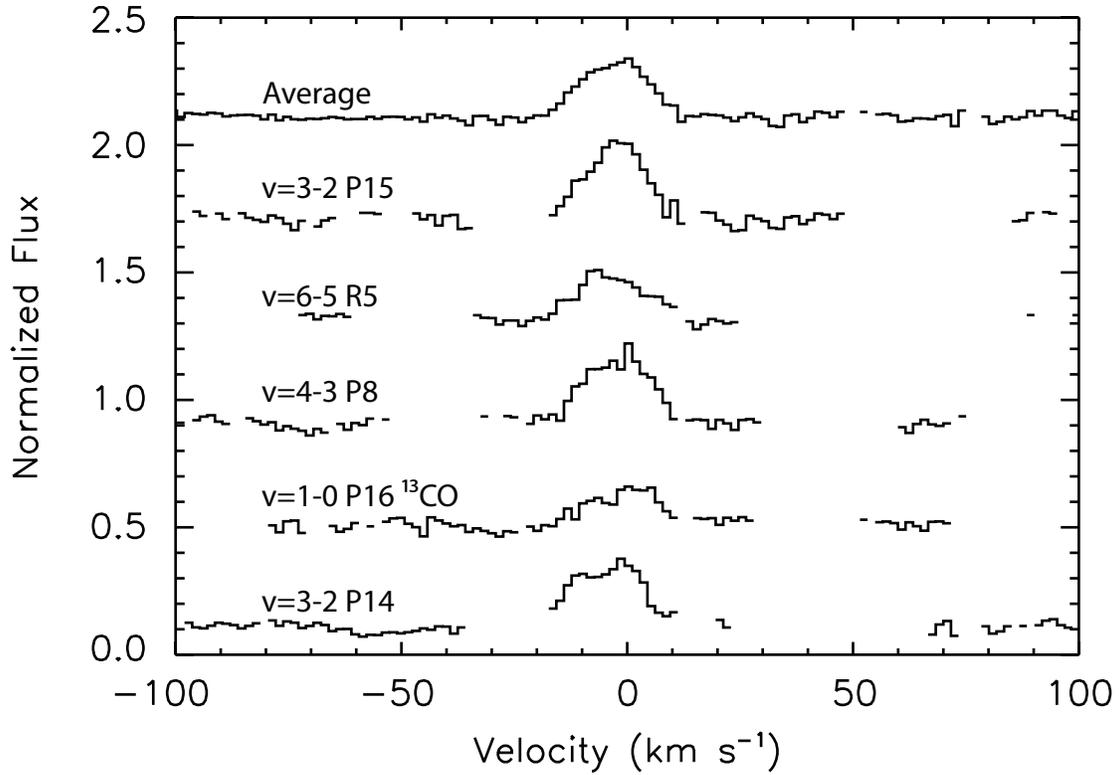}
	\caption[CO lines]{Unblended hot band CO lines in the rest frame of the star. The signal to noise of the continuum of the average profile is $\sim$110. The HWZI of the line is 13~$\pm2$~km~s$^{-1}$. Line positions are reported in Table 3.}	
	\label{fig:obs6}
   \end{center}
\end{figure}

Because the individual OH lines are of low signal to noise (a result of their low equivalent width) we constructed an average profile of the unblended OH lines in order to facilitate the comparison of the OH, CO, and [\ion{O}{1}] $\lambda$6300\AA\ line profiles (Fig. \ref{fig:obs5} and Fig. \ref{fig:obs6}).  

\begin{figure}
    \begin{center}
	\includegraphics[scale=1.]{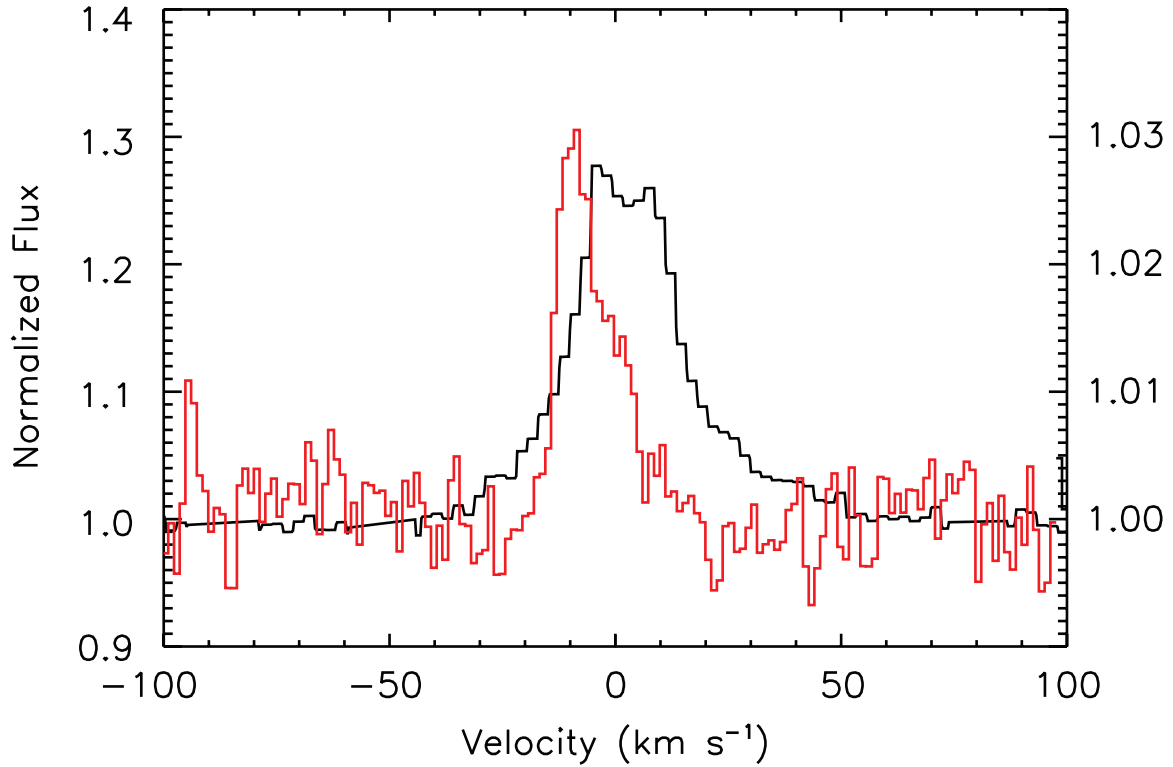}
	\caption[OI comp]{Comparison of the [\ion{O}{1}] ~$\lambda$6300\AA\ line and the average OH line profile (adapted from \citet{2005A&A...436..209A}) in the rest frame of the star. 
	The scale of the [\ion{O}{1}] $\lambda$6300\AA\ line (black) is the left vertical axis.  The scale of the OH emission line (red) is denoted on the right vertical axis. The line profiles are clearly different, and the HWZI velocity of the OH emission is much narrower than that of the \ion{O}{1} line indicating that it does not arise from the same region of the disk as the OH emission.}
	\label{fig:obs7}
   \end{center}
\end{figure}

\begin{figure}
    \begin{center}
	\includegraphics[scale=1.]{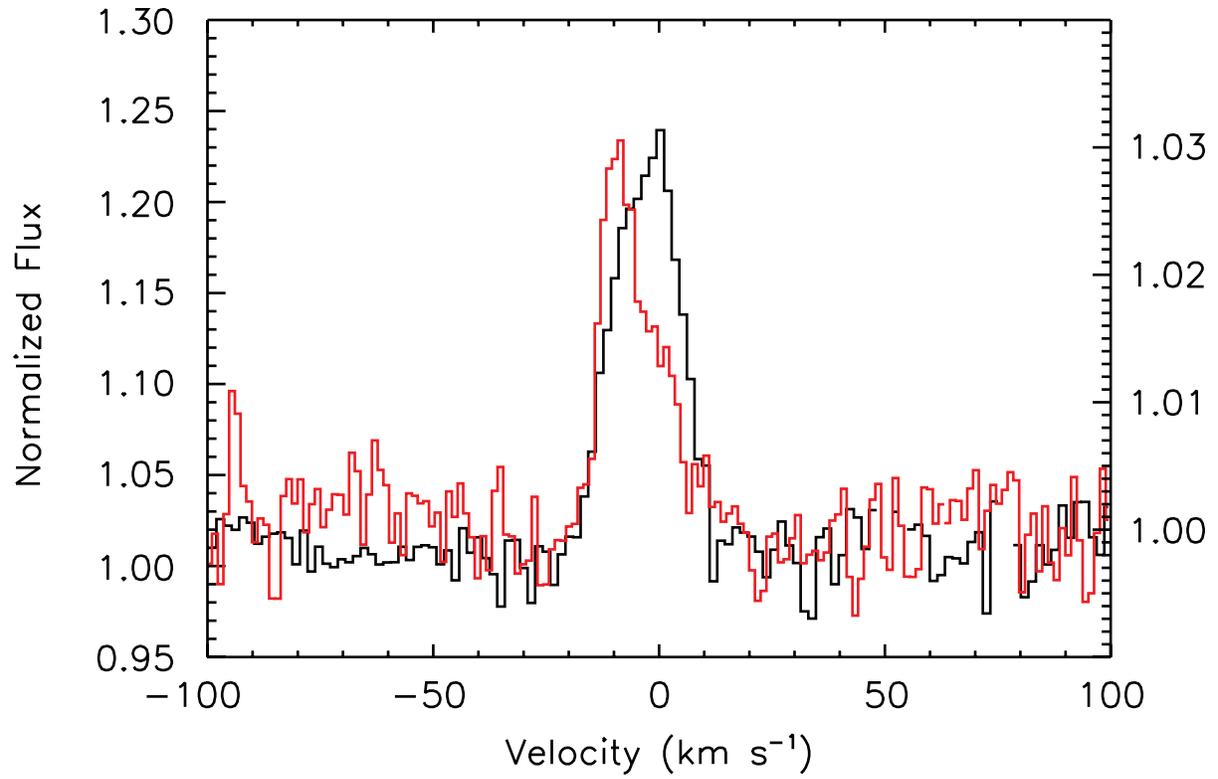}
	\caption[OHCO comp]{Comparison of the average hot band CO line profile (black) and the average OH line profile (red) in the rest frame of the star. The scale of each emission line profile is plotted in the left axis. The line profiles are clearly different, but the width of each line is consistent.} 
	\label{fig:obs8}
   \end{center}
\end{figure}

The full width at zero intensity (FWZI) of the OH emission is 24~km~s$^{-1}$ (Fig. \ref{fig:obs7}) and is consistent with the CO emission (Fig. \ref{fig:obs8}). The [\ion{O}{1}]~$\lambda$6300\AA\ emission line (Fig. \ref{fig:obs7}) has a FWZI of 90~km~s$^{-1}$. While the \ion{O}{1} extends to about 0.8~AU \citep{2005A&A...436..209A}, the inner extent of the OH is consistent with that of the CO which is truncated at 13~$\pm$6~AU \citep{2009ApJ...702...85B}.

The profiles of the OH lines differ from the profiles of the hot band CO lines. Inspection of the average CO line profile reveals a modest asymmetry where the emission peaks on red side of the line, though the blue side contains the majority of the flux. This is similar to what was observed in 2006 \citep{2009ApJ...702...85B}. The OH line profile is also resolved and asymmetric (Figs. \ref{fig:obs5}, \ref{fig:obs7}, \ref{fig:obs8}). The flux ratio of the blue side of the OH line to the red side of the OH line is $\sim$4.

We determine the Doppler shift of these asymmetric lines by fitting our model to the data. The barycentric Doppler shift is +16~km~s$^{-1}$. This is similar to the radial velocity we infer from submillimeter observations of CO \citep[$\sim$14~km~s$^{-1}$;][]{2010A&A...519A.110P}. It is also similar to the value inferred by observation of the [\ion{O}{1}] $\lambda$6300\AA\ line \citep[$\sim$18~km~s$^{-1}$;][]{2005A&A...436..209A}.

\subsection{Origins of the Asymmetry}

We consider four possibilities for the origin of the asymmetry seen in the average OH line.

Firstly, a wind may generate the blue-shifted asymmetry seen in the OH line.  \citet{2009ApJ...702..724P} show that the [\ion{Ne}{2}] 12.81 $\mu$m emission in multiple T Tauri stars has a blue-shifted centroid of approximately a few~km~s$^{-1}$ and a slight blue excess that they attribute to a disk wind.  One difference between the [\ion{Ne}{2}] emission and the OH line profile studied here is that the [\ion{Ne}{2}] line profiles are marginally asymmetric, whereas our OH line profile is dramatically so. 

Molecular line profiles indicative of disk winds have also been reported, although the line profiles are also much more symmetric than our OH line profile.  In their study of CO emission from T Tauri stars, Bast et al. (2011) reported symmetric CO line profiles with a narrow peak and broad wings for a subset of the sources they studied. The spatial constraints on the emission led the authors to conclude that the emission likely arises from a combination of emission from a circumstellar disk and a slow-moving disk wind. Detailed modeling by Pontoppidan et al. (2011) shows that such a combination can produce a reasonable fit to the observed line profiles and their spectroastrometric properties. Additionally, NIR emission from organic molecules has been detected for several T Tauri stars, with line profiles similar to that found for the CO emission, suggesting a similar origin (Mandell et al. 2012). For systems with large disk inclinations like that in HD~100546, self-absorption is expected if the wind component is strong, which is not seen in HD~100546  (i=$45^{\circ}$). Ro-vibrational CO emission lines forming in outflowing material of V1647 Ori have also been observed, but these have a P~Cygni profile and a symmetric emission component  \citep{2007ApJ...670L..29B}, a different line profile from that of the OH in HD~100546. While we cannot rule out that the ro-vibrational OH lines we observe form in a disk wind that gives rise to a blue-shifted excess, our profile is distinct from the other profiles in the literature attributed to a disk wind.

Secondly, transonic turbulence in the disk (e.g., driven by the magnetorotational instability) produces spatial inhomogeneities in the temperature and density distribution of the disk. If they are large, such inhomogeneities might produce an OH line profile as asymmetric as observed.  Simulations by \citet{2006A&A...457..343F} find 1$\sigma$ fluctuations from the mean density of $\sim$10\%. A much larger fluctuation is likely necessary to produce the level of asymmetry that we observe. An additional potential difficulty with this picture is that we observe a much more dramatic asymmetry in the OH line profile than the CO profile. Could both profiles arise from the same turbulent disk?  If the critical densities of CO and OH for ro-vibrational excitation differ significantly, their sensitivity to inhomogeneities might be more pronounced for one molecule than the other.  This picture could be tested with future observations.  If the asymmetry originates from turbulence in the disk, the OH line profile would be expected to show stochastic line profile variations, and we would not continue to see the profile that we now observe.

Thirdly, a localized hot-spot or a dense lump of material could induce non-axisymmetric emission.  Such a hot-spot in the disk could be caused by a planet/disk interaction if the planet warms the inner rim in its vicinity. If this is the case, then the profile of the line should vary with the orbital phase of the companion.

Finally, hydrodynamic modeling indicates that a gas giant planet in a circumstellar disk can induce an eccentricity as high as 0.25 in the inner rim that falls off as $\rm \sim r^{-2}$ \citet{2006A&A...447.396K}.  The modeling also predicts that the semi-major axis of the eccentric disk precesses slowly ($\sim$10$\degr /1000$ orbits). Thus the OH line profiles should show minimal variation over the orbital period of the planet. To determine whether an eccentric inner rim can account for the observed asymmetry, we model the hot band CO and OH emission line profiles assuming that the inner wall has a nonzero eccentricity while the outer disk is circular.

 \section{Modeling}
 \label{sec:Modeling}

 In our model, we adopt the disk geometry described by \citet{2009ApJ...702...85B}. The inner rim is puffed up \citep{2001ApJ...560..957D} with a scale height of 3.5AU \citep{2003A&A...401..577B, 2009ApJ...702...85B, 2010A&A...511A..75B}. We adopt the inner radius found from the synthesis of the ro-vibrational CO emission \citep[13~AU;][]{2009ApJ...702...85B} and the same radial profile for the disk emission found from our previous modeling ($\sim$R$^{-1/2}$).  We use a stellar mass of 2.4 $M_{\odot}$ \citep{1997A&A...324.33V}, and an inclination of $45 \degr $, a value within the bounds measured by \citet{2011ApJ...738...23Q} ($47 \degr \pm 2.7 \degr $) and \citet{2007arXiv0704.1507A} ($42 \degr \pm 5 \degr $).
 
\begin{figure}
    \begin{center}
	\includegraphics[scale=.7]{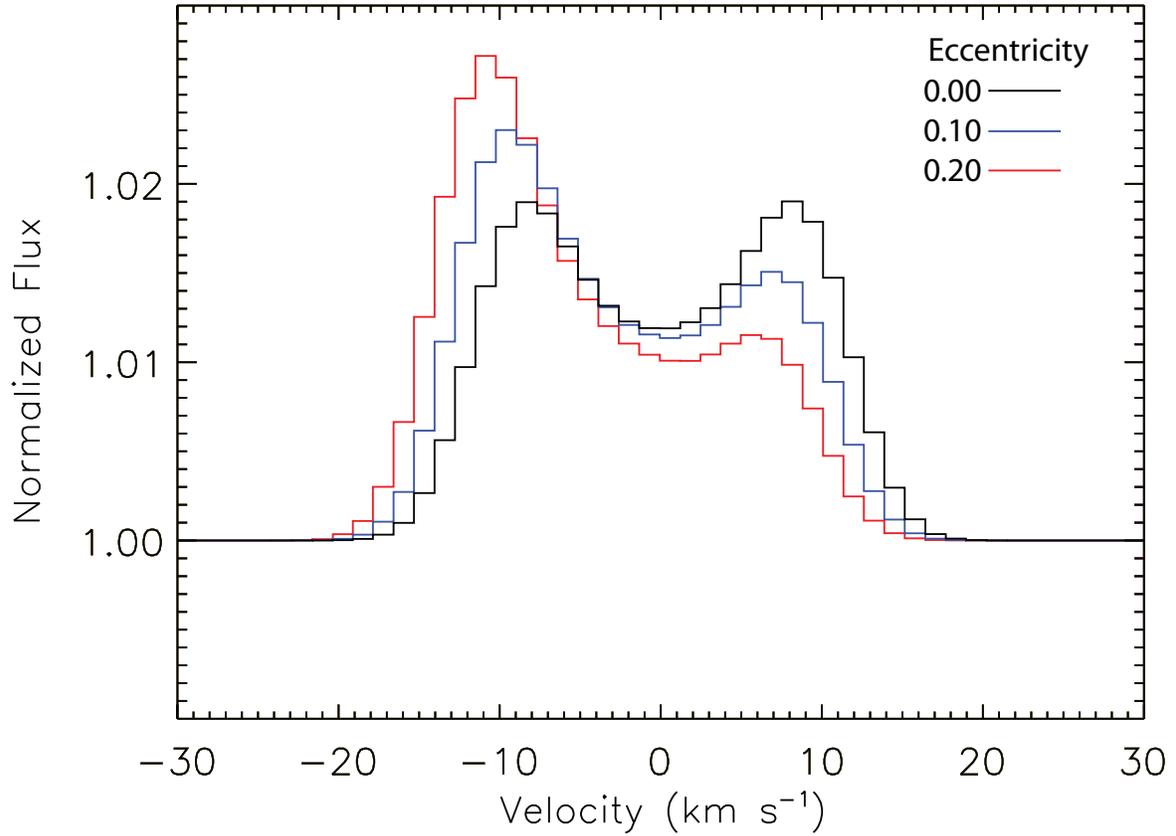}
	
	\caption[Synthetic Lines]{Synthetic, spatially unresolved line profiles of emission from an eccentric inner wall that is inclined by 45\degr to the line-of-sight in the rest frame of the star. The other model parameters are alpha =-1, an radius for the inner wall of 13 AU, and a stellar mass of $2.4 M_\odot$ (see text for details). The emission is entirely from the wall component in this example to best illustrate the asymmetry, which grows as we increase the eccentricity of the wall.  For this example the equivalent width is held constant and equal to the measured equivalent width of the average OH line. The asymmetric emission bump is blueward of line-center due to the direction we have chosen for the rotation of the gas and the viewing angle of disk; these choices are entirely general.  For HD~100546 we see a similar blueward emission bump in the OH emission (see text).}
	
	\label{fig:obs9}
   \end{center}
\end{figure}

 The magnitude of the bulk velocity of a parcel of gas in an eccentric orbit is not axisymmetric and is given by $v(\theta)^2=GM_\star(1+e^2 + 2ecos(\theta))/(a(1-e^2))$, where $G$ is the universal gravitational constant, $M_\star$ is the stellar mass, $a$ is the semi-major axis of the eccentric annulus, $e$ is the eccentricity, and $\theta$ is the phase of the orbit measured from periastron. For non-zero eccentricities, the velocity of the gas at periastron is larger than at apastron, so the side of the line profile arising from gas near periastron will extend to higher velocities (Fig. 9). Additionally, the gas in the disk near periastron is closer to the star making that gas warmer and brighter. We assume the temperature scales as $T = T_{0}(R/R_{0})^{\alpha}$ and the luminosity of the line is proportional to $ B_{\tilde{\nu}}$$(1 - e^{-\tau_{\tilde{\nu}}})$. 
Fig. \ref{fig:obs9} shows the effect of increasing eccentricity on a line profile. The synthetic emission line profiles shown arise from an eccentric, unresolved annulus inclined by 45\degr with $\alpha = -1.$

In modeling the line profiles, we assume the line of sight falls along the semi-minor axis of the disk and the slit is aligned with the semi-major axis of the disk.\footnote{The position angle of a disk is defined by the apparent semi-major axis of the inclined disk. The apparent semi-major axis is not generally aligned with the intrinsic semi-major axis of the eccentric disk. In this paper the semi-major and semi-minor axes refer to intrinsic axes of the disk not the apparent axes.} Such an orientation maximizes the difference between the projected velocity on the red-shifted and blue-shifted sides of the disk. If the line of sight falls along the semi-major axis, this results in the gas at apastron and periastron having a zero projected velocity and a symmetric line. Thus the eccentricity inferred from our modeling is a lower limit. 
 
For the line profile calculation, the disk is divided into annuli, and each annulus is then further divided azimuthally into zones, with each representing a change in the projected velocity (along the line of sight) of 1~km~s$^{-1}$. We assume that the inner wall has an eccentricity that is left as a free parameter. The annuli comprising the surface of the disk are assumed to have an eccentricity of zero.  
In the expression for the radial temperature variation, $T = T_{0}(R/R_{0})^{\alpha}$, the fiducial temperature $T_{o}$ is assumed to be 1400~K based on what was observed for CO.  The contribution to the luminosity of the emission lines from each zone is proportional to the area of the zone and $ B_{\tilde{\nu}}$$(1 - e^{-\tau_{\tilde{\nu}}})$. The emergent spectrum is then calculated by looping over all the zones and shifting the luminosity arising from each zone by the projected velocity.  The spectrum is then convolved with the instrument profile, scaled to the 
flux of the OH feature, and resampled into a common velocity space.  We leave the radial dependence on temperature (and thus the luminosity) of the OH emission in the inner rim a free parameter.
 
The fraction of the emission that arises from the inner wall of the disk is also left as a free parameter for the OH and CO distributions. We achieve the best fit if 75$\%$ of the OH arises from the inner edge, which is similar to what has been found for CH$^+$ in HD~100546 \citep{2011A&A...530L...2T}. The remaining flux comes from the outer disk. For the best fit to the CO, $35\%$ of the luminosity arises from the wall and 65$\%$ from the surface of the disk.  Fluorescence modeling in \citet{2009ApJ...702...85B} found a slightly smaller percentage $(15\%)$ of the luminosity arises from the inner wall. This ratio can be changed in their model by increasing the turbulent velocity of the gas in the inner rim and decreasing the flaring or turbulent velocity of the outer disk.  

\begin{figure}
    \begin{center}
	\includegraphics[scale=1.]{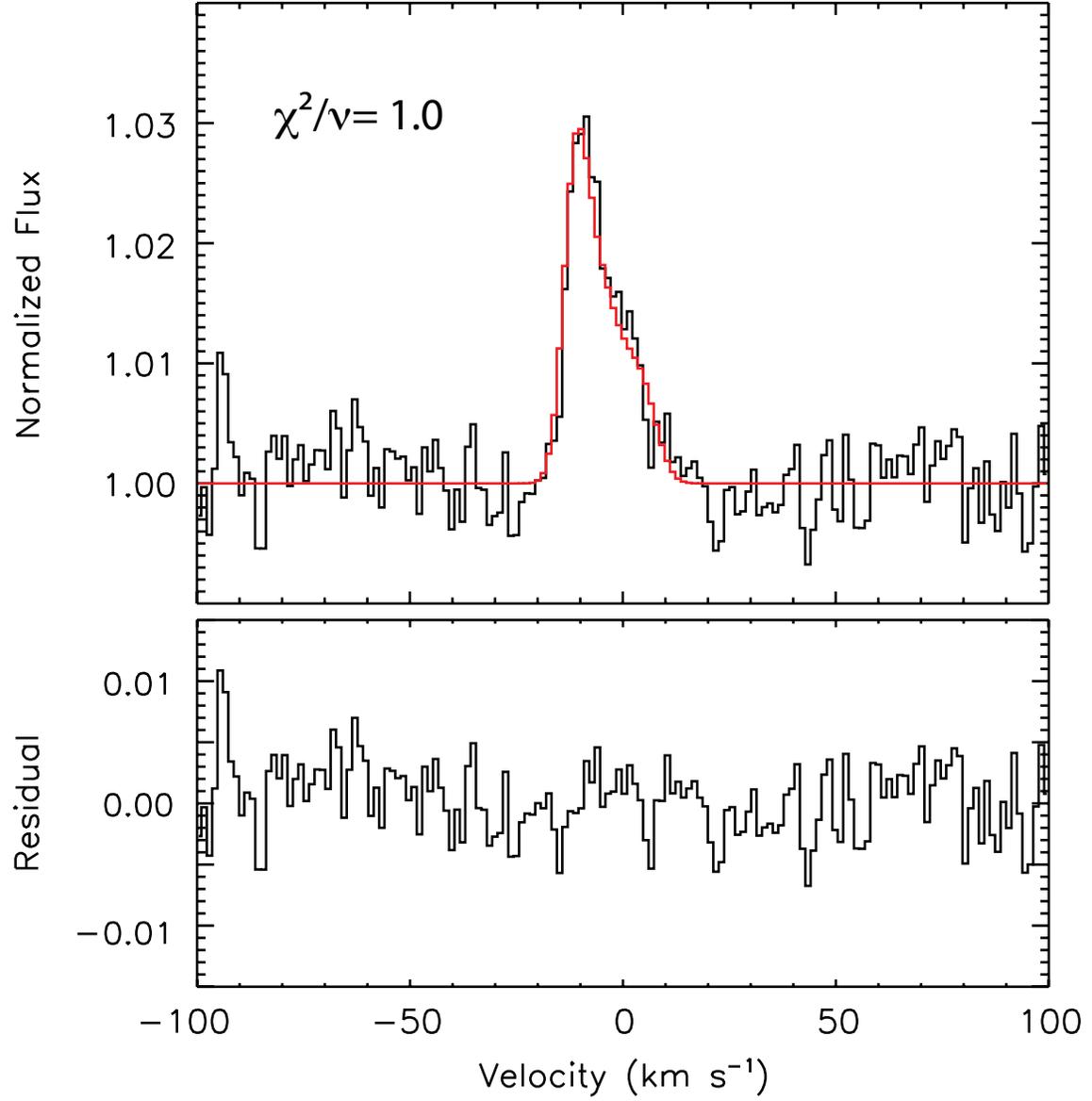}
	\caption[model fit]{Synthesized spectrum of OH emission arising from the inner wall of a disk with an eccentricity of 0.18 in the rest frame of the star (red line).  The average OH line profile (black line) is plotted in the upper panel.  In the lower panel we plot the difference between the data and the model.  To reproduce these line shapes we assumed that the ratio of the emission luminosity of the wall to the luminosity of the disk was approximately 3:1.  Lower ratios produce a more double-peaked structure, while higher ratios do not adequately fill out the red component of the OH emission line.  The minimum reduced chi-squared value, $\chi^{2}/\nu$, is 1.0.}  
	\label{fig:obs10}
   \end{center}
\end{figure}

\begin{figure}
    \begin{center}
	\includegraphics[scale=1.]{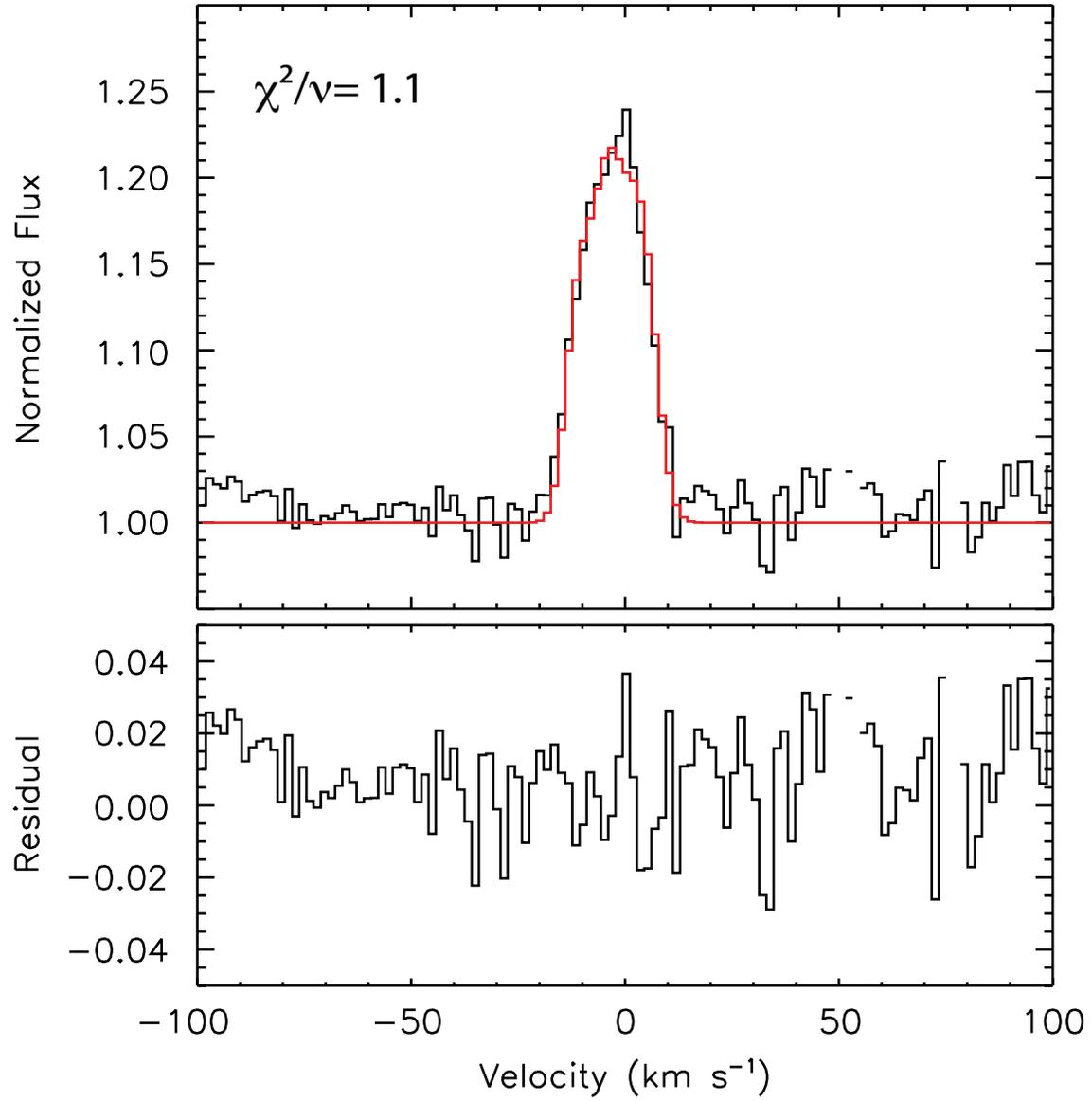}
	\caption[model fit]{Synthesized spectrum of CO emission arising from the inner wall and the disk in the rest frame of the star (red line).  The average line profile of the unblended hotband CO emission lines (black line) is plotted in the upper panel. The difference between the observed line profile and the modeled line profile is plotted in the lower panel.  The synthetic CO line profile includes components from both the inner wall and disk.  We assumed that the ratio of the emission luminosity of the wall to the luminosity of the disk was approximately 1:3.   The reduced chi-squared value, $\chi^{2}/\nu$, is 1.1.}
	\label{fig:obs11}
   \end{center}
\end{figure}

We are able to fit the line profiles with the following values (and associated $99.7\%$ confidence levels): $e=0.18$ (0.07 to 0.30), $\alpha= -2.5$ $(< -1.2)$, an OH fraction (arising from the inner wall) of 0.75 ($>0.44$), and a CO fraction (arising from the inner wall) of 0.35 (0.19 to 0.55).  The uncertainty in the eccentricity does not include the uncertainty in the viewing geometry.  The minimum reduced $\chi^{2}$ for fits corresponding to a sampled parameter space near these values is 1.0, which corresponds to the overall reduced $\chi^{2}$ of the model fits shown in Fig. \ref{fig:obs10} \& Fig. \ref{fig:obs11}. If the line of sight does not fall along the semi-minor axis of the disk, then the eccentricity of the disk will be higher.  If the line of sight to the disk is offset from the semi-minor axis by 10$\degr$, 20$\degr$, and 30$\degr$, the eccentricity that results in the best fit for each line of sight is 0.28 , 0.45, and 0.48, respectively.  
 
The transition from the eccentric inner wall to the circular outer disk is dynamically complicated, but models of such disks indicate the transition is sharp \citep{2006A&A...447.396K,2010A&A...523A..69R}. Gas emitting from the surface of the disk within a few AU of the inner rim makes a minimal contribution to the line profile. As a result, the lack of smooth continuity between the eccentric inner annulus and the rest of the circular disk does not appreciably affect our line profile.

 \section{Discussion}
 \label{sec:Discussion}

\subsection{Massive Companion}

One possibility for the origin of the eccentricity of the inner disk is tidal interactions with an embedded gas giant planet.  When a gas giant planet forms in a disk, tidal interactions between the planet and the disk cause the inner edge of the outer disk to grow eccentric \citep{2006A&A...447.396K,2001A&A...366..263P,1991ApJ...381..259L}. In these hydrodynamic simulations, the 1:3 Lindblad resonance induces the eccentricity. The 1:2 Lindblad resonance has a damping effect on the generation of the eccentric structure. Thus the gas must be cleared from this region of the disk.

For companion masses greater than q=0.003, where q is the ratio of the companion to stellar mass, and for a viscosity parameter value of $\alpha$ $\sim$.004, which is typical of circumstellar disks \citep{1999MNRAS.303..696K}, \citet{2006A&A...447.396K} show that significant eccentricities (e $\lesssim$ 0.25) can be induced in the portion of the disk exterior to the orbit of the embedded gas giant. The azimuthally averaged eccentricity falls off approximately as $r^{-2}$ where $r$ is the distance from the star. As a result,  only a narrow annulus of the disk has significant eccentricity (the precise radial dependence of the eccentricity depends on the viscosity of the gas).

The possibility of a massive companion to HD~100546 has been explored previously by several authors. Firstly, the transitional nature of the SED may be indicative of  dynamical sculpting of the disk by a massive companion (e.g. \citet{skrutskie1990, 1992ApJ...395L.115M,1999ApJ...514..344B, 2002ApJ...568.1008C, 2003MNRAS.346L..36R,  2004ApJ...612L.137Q, 2005ApJ...621..461D, 2005ApJ...630L.185C, 2006ApJS..165..568F}. Indeed,  \citet{2011arXiv1110.3808K} have presented intriguing imagery of the transitional object Lk~Ca~15 consistent with the presence of a forming planet. \citet{2003A&A...401..577B}  model the SED of HD~100546 and conclude that the inner $\sim$10~AU of the disk is cleared. These authors suggest that the companion must be at least 5.6~M$\rm_J$ in order to open the hole in the disk.  Secondly, \citet{2005ApJ...620...470G} used long slit images acquired with the $Space\ Telescope\ Imaging\ Spectrograph$ on the $Hubble\ Space\ Telescope$ and confirm that the inner edge of the disk is at 13$~\pm$3AU. They also find that the star is not centered in the inner hole of the disk. Rather the star falls 5$~\pm$3~AU to the northwest. If the offset is due to the eccentricity of the disk, then $e=0.38\pm0.24$. This eccentricity is consistent with our result from modeling the line profile of the OH emission. In Paper II, we discuss how the observed variability in the CO fundamental emission may be an additional signpost of a forming gas giant planet orbiting HD~100546. 

The result of our spectral synthesis of the asymmetric OH line profile is similar in spirit to the predictions of \citet{2010A&A...523A..69R}, although the observations differ in detail from their model predictions.   \citet{2010A&A...523A..69R} carried out hydrodynamic simulations of a disk with an embedded planet and used the ro-vibrational CO emission as a tracer in exploring how disk emission line profiles are modified by the presence of the planet.  They confirmed the finding of \citet{2006A&A...447.396K} that an embedded planet induces a significant eccentricity in the disk.  Planets above a critical mass ($\geq$3 M$\rm_{J}$ for a 1 M$_\sun$ star) will cause the outer edge of the gap opened by the planet to be elliptical in shape and induce a measurable asymmetry in the CO line profiles. Based on our line profile modeling, we anticipate detecting a small but observable velocity shift in the line profiles due to the difference between the projected velocities at apastron and periastron. 

In the \citet{2010A&A...523A..69R} predictions for CO emission, the disk extends from $\sim$0.2~AU to 5 AU, and CO emits over this region of the disk, i.e., within, at, and beyond the orbit of the planet.  In comparison, we do not measure any emission from the inner optically thin region of the disk around HD~100546 (i.e., within 13 AU).  Instead, the CO lines excited by UV fluorescence arise from the eccentric inner rim and the circular outer disk extending from $\sim$13-100 AU. The OH emission is mostly restricted to the inner rim of the disk. Thus the CO and OH line profiles of HD~100546 appear different from the synthetic profiles generated by \citet{2010A&A...523A..69R}, but several of the elements of their scenario are consistent with the observations. 
 
 \subsection{Comparison of OH and [OI] Line Profiles}

Our results complement other studies of OH disk emission in the literature.  In their study, \citet{Fedele2011} detected only the P4.5 line and therefore could not determine a temperature or the excitation mechanism with the available data.  The authors do show that the  [\ion{O}{1}] $\lambda$6300\AA\ line and OH emission lines observed toward HD~259431 have the same profile, providing some support to the hypothesis that the  [\ion{O}{1}] $\lambda$6300\AA\ line arises from the photodestruction of OH. They also find that OH emission is more common from flared disks, which they suggest indicates the importance of fluorescent excitation.  CO observations by \citet{2007ApJ...659..685B} further show the width of the P30 ro-vibrational line to be similar to that observed in the OH and  [\ion{O}{1}] line for HD~295431.
  
\citet{Fedele2011} also observed asymmetric ro-vibrational OH lines toward V380 Ori, which they suggest could be caused by a deviation from axisymmetric Keplerian rotation or by a non-homogenous distribution of gas.  Curiously, \citet{2009MNRAS.400..354A} find that V380 is a spectroscopic binary with a separation of rsin(i)$\lesssim$0.33~AU.  Asymmetric ro-vibrational CO lines similar to the OH ro-vibrational lines presented here were observed in the recent outburst of EX Lupi \citep{2011ApJ...728....5G}.  It should be noted that the EX Lupi system is highly variable due to rapid accretion; $\dot{M} \sim 10^{-7}~M_{\odot}~yr^{-1}$, \citep{2010ApJ...719L..50A}. Direct comparison to line formation regions in HD~100546 is problematic due to HD~100546's comparably slow accretion rate; $\dot{M} \sim 10^{-9}~M_{\odot}~yr^{-1}$, \citep{2005ApJ...620...470G}.

The [\ion{O}{1}] $\lambda6300$\AA\ emission from HD~100546 has been interpreted as arising from the photodissociation of OH \citep{2006A&A...449..267A}.  \citet{1998ApJ...502L..71S} show that 
in H/H$_{2}$ photodissociation regions far ultraviolet photons dissociate the OH molecule leaving just over half of the oxygen atoms in the $\rm^1D_2$ state - the upper level of the [\ion{O}{1}]~$\lambda6300$\AA\ emission line. The first pre-dissociation band of CO is near 1100\AA where the UV field of HD~100546 is quite weak and atomic hydrogen can provide some shielding. In contrast, the first pre-dissociation band of OH is centered near 1600\AA~\citep[$1^2\Sigma^-$;][]{1984ApJ...277..576V}, which falls near the 4th positive system of CO. Thus CO and OH can self-shield one another.  \citet{2006A&A...449..267A} model the line profile of the \ion{O}{1} emission from HD~100546 and show the gas is distributed from 0.8-100~AU. Building on the work by \citet{2000ApJ...539..751S}, they find that the surface density of the OH necessary to give rise to the \ion{O}{1} emission is,
\begin{equation}
\Sigma_{OH}(R)=3.6\times10^{21}(R/AU)^{-2.5}.
\end{equation}
This  interpretation requires the inner disk to be curiously rich in OH and very depleted in CO (N(OH)/N(CO)$\gg$100)â contrary to predictions for the chemistry of inner disks \citep[e.g.][]{thi05}.

Adopting this OH surface density and assuming the gas is in LTE, we find that the OH would need to be as cool as 450~K or have a low density to fall below our detection limit. Our upper limit on the rotational temperature of the OH is 1400~K which is comparable to the temperature of the CO at the inner rim of the disk (1400$\pm$400~K; \citet{2009ApJ...702...85B}). It is not clear why OH in the inner disk would be significantly cooler than CO beyond 13 AU. In principle, it is also possible that the gas in the inner disk is below the critical density necessary to thermalize the vibrational levels such that the effective vibrational temperature is much less than 1400~K. However, electronic excitation of the OH molecule can also populate the excited vibrational levels. If the OH is not electronically excited because it is shielded from the NUV radiation field, then it is unlikely that a significant quantity of OH is photodissociated and able to give rise to the observed \ion{O}{1} emission. Thus it seems more likely that the inner disk is largely devoid of OH as well as CO and the [\ion{O}{1}]~$\lambda$6300\AA\ line likely arises from a process other than the photodestruction of OH.  Detailed modeling of the excitation of the OH will be considered in future work.

\section{Conclusions}
\label{sec:conclusions}
We have shown that the observed asymmetric emission line profile of OH can be reproduced as arising from an inner wall with an eccentricity that is in line with current estimates of the eccentricity induced by an orbiting gas giant companion. While we cannot conclusively demonstrate that the emission arises from an eccentric rim, our hypothesis is testable. If the asymmetry is due to the eccentricity of the inner rim induced by a massive companion, then the line profile should remain approximately constant in time. One caveat is that if the orbiting planet warms the inner rim of the disk in its vicinity, the emission line profiles may shift position in velocity space as the planet moves through its orbit. If confirmed, this observation provides additional circumstantial evidence that the gap in the disk at 13~AU has been opened by a massive companion. 

Other possible explanations for the line asymmetry include an origin in a wind or a turbulent disk atmosphere.  If the asymmetry is a consequence of disk turbulence, the profile shape should vary stochastically with time, and we would not continue to see the profile that we now observe.  If the profile arises in a wind, the asymmetry should also be present in other molecular line diagnostics that probe similar physical conditions.

We have also shown that the OH emission lines are significantly narrower than the [\ion{O}{1}]~$\lambda6300$\AA\ emission line. Thus we conclude that the OH and \ion{O}{1}  are unlikely to be co-spatial. The lack of OH emission from the inner disk suggests that the destruction of OH is not responsible for the [\ion{O}{1}]~$\lambda$6300\AA\ emission line, which motivates further study into the nature and origin of the emission.

\acknowledgments
Based on observations obtained at the Gemini Observatory, which is operated by the Association of Universities for Research in Astronomy, Inc., under a cooperative agreement with the NSF on behalf of the Gemini partnership: the National Science Foundation (United States), the Particle Physics and Astronomy Research Council (United Kingdom), the National Research Council (Canada), CONICYT (Chile), the Australian Research Council (Australia), CNPq (Brazil) and CONICET (Argentina). The Phoenix infrared spectrograph was developed and is operated by the National Optical Astronomy Observatory.  The Phoenix spectra were obtained as part of programs GS-2010B-C-2. We would like to thank Bram Acke for providing the published spectrum of \ion{O}{1} for comparison with our OH spectrum. S.D.B. and J.P.L. acknowledge support for this work from the National Science Foundation under grant number AST-0708899 and NASA Origins of Solar Systems under grant number NNX08AH90G. Basic research in infrared astronomy at the Naval Research Laboratory is supported by 6.1 base funding.

\appendix
\section{Appendix A: Spectroscopic Notation}
The total rotational angular momentum is given by the number following the P, the numerical subscript corresponds to electron spin of either $-1/2$ (1) or $+1/2$ (2), and the $+/-$ signs correspond to even/odd parity. 
 \linebreak
 
Unlike common diatomic molecules such as $H_{2}$ and CO, the total orbital angular moment of the ground electronic state of the OH radical is $\Lambda$=1 and is thus doubly-degenerate. The additional degrees of freedom require a somewhat more extensive labeling scheme to fully describe the transitions. In this paper, we adopt the nomenclature described by \citet{1982ApJ...258..864S}. Due to the interaction of the electron spin and orbital angular momentum (spin-orbit coupling), the ground electronic state is split into $^2\Pi_{\Lambda-1/2}$ and $^2\Pi_{\Lambda+1/2}$. The total angular momentum, $J$, is given by the sum of the molecular rotation, $N=0, 1, 2, 3, ...$ and electron spin $s=\pm1/2$, so that $J=N+1/2$ in the $X^2\Pi_{3/2}$ state and $J=N-1/2$ in the $X^2\Pi_{1/2}$ state. In addition, each rotational level is split into two states due to the coupling of the molecular rotation to the orbital motion of the electrons (lambda-type doubling); these split levels are labeled as $(+)$ or $(-)$ based on the parity of the level. The selection rules for a ro-vibrational transition in the ground electronic state are $\Delta J=0,\pm1$ and $\pm \Rightarrow \mp$. 
\linebreak

Rotational transitions of $J^{\prime \prime}=J^\prime-1$,  $J^{\prime \prime}=J^\prime $, and $J^{\prime \prime}=J^\prime+1$ are labeled $R,~Q,$ and $P$ respectively. For transitions with $N+1/2$ (i.e. the $^2\Pi_{3/2}$ ladder), the rotational designation is labeled with the subscript $2$. For $N-1/2$ the rotational designation is labeled with the subscript $1$. When the molecule undergoes a transition between the $^2\Pi_{3/2}$ ladder to the $^2\Pi_{1/2}$ ladder, the 1 and 2 are given in the order of the transition (upper level first). The parity of the ground state is then labeled with the superscript $+$ (for even parity) or $-$ (for odd parity). The ground state electronic level is labeled $X$, the first excited bound electronic level $A$, the second excited bound electronic level $B$, etc. In this paper we are only concerned with ro-vibrational transitions within the ground electronic state and thus leave off the $X$.

\clearpage

\begin{deluxetable}{lllcccc}
\tablenum{1}
\tablewidth{0pt}
\tablecaption{Log of Observations}

\tablehead{ \colhead{Date} &\colhead{Telescope/Instrument} & \colhead{Target}& \colhead{Spectral Grasp}  &  \colhead{Integration}  & \colhead{S/N} & \colhead{Airmass}\\
\colhead{}  & \colhead{} & \colhead{} & \colhead{cm$^{-1}$} & \colhead{minutes}  & \colhead{} & \colhead{}} 
\startdata
2010 Dec 22 & Gemini S./PHOENIX & Theta Car & 3138 - 3150 & 8  & - & 1.5\\
2010 Dec 22 & Gemini S./PHOENIX & HD~100546 & 3138 - 3150 & 40  & 180 & 1.6 \\
2010 Dec 23 & Gemini S./PHOENIX & Theta Car & 2837 - 2850 & 8  & - & 1.7 \\
2010 Dec 23 & Gemini S./PHOENIX & Theta Car & 2027 - 2037.5 & 8 & - & 1.4 \\
2010 Dec 23 & Gemini S./PHOENIX & HD~100546 & 2837 - 2850 & 28  & 180 & 1.8\\
2010 Dec 23 & Gemini S./PHOENIX & HD~100546 & 2027 - 2037.5 & 32 & 79 & 1.4 \\
\enddata
\label{tab:observations}
\end{deluxetable}

\begin{deluxetable}{lcc}
\tablenum{2}
\tablewidth{0pt}
\tablecaption{Equivalent Widths of OH lines}

\tablehead{ \colhead{Line} & \colhead{Wavenumber} & \colhead{Equivalent Width}\\
\colhead{} & \colhead{cm$^{-1}$} & \colhead{$\rm 10^{-3}cm^{-1}$}}
\startdata

P9.5$_{2^-}$ 	& 3146.18 & 3.8$\pm$0.5\\
P9.5$_{2^+}$	& 3145.49 & 4.1$\pm$0.5\\
P10.5$_{1^+}$	& 3142.06	 & 3.8$\pm$0.5\\	  
P10.5$_{1^-}$	& 3141.04	 & 5.3$\pm$0.5\\
P15.5$_{2^-}$ & 2840.82 & $<$.5\\
P15.5$_{2^+}$ & 2839.58 & $<$.5\\
P16.5$_{1^+}$ & 2839.03 & $<$.5\\
P16.5$_{1^-}$ & 2837.43 & $<$.5\\

\enddata
\label{tab:EqWidths}
\end{deluxetable}

\begin{deluxetable}{lc}
\tablenum{3}
\tablewidth{0pt}
\tablecaption{Unblended Lines}
\tablehead{ \colhead{Line} & \colhead{Wavenumber}\\
\colhead{} & \colhead{cm$^{-1}$}}
\startdata

v=3-2 P15 & 2030.16 \\
v=6-5 R5 & 2032.83 \\
v=4-3 P8 & 2033.14 \\ 
v=1-0 P16 $^{13}$CO & 2033.42 \\
v=3-2 P14	 & 2034.41 \\

\enddata
\label{tab:cleanlines}
\end{deluxetable}%

\end{document}